# Facile Self-Assembly of Quantum Plasmonic Circuit Components

*Toan Trong Tran,[1] Jinghua Fang,[2] Hao Zhang,[1] Patrik Rath,[3] Kerem Bray,[1] Russell Sandstrom,[1] Olga Shimoni,[1] Milos Toth,[1,*] and Igor Aharonovich[1,*]*

1. School of Physics and Advanced Materials, University of Technology Sydney, Ultimo, NSW 2007, Australia
*E-mail: Igor.Aharonovich@uts.edu.au; Milos.Toth@uts.edu.au

2. Plasma Nanoscience Laboratories, Manufacturing Flagship, Commonwealth Scientific and Industrial Research Organisation (CSIRO), P.O. Box 218, Lindfield, NSW 2070, Australia

3. Institute of Nanotechnology, Karlsruhe Institute of Technology, 76344 Eggenstein-Leopoldshafen, Germany

Efficient coupling between solid state quantum emitters and plasmonic waveguides is important for the realization of integrated circuits for quantum information, communication and sensing. However, realization of plasmonic circuits is still scarce, particularly due to challenges associated with accurate positioning of quantum emitters near plasmonic resonators. Current pathways for the construction of plasmonic circuits involve cumbersome and costly methods such as scanning atomic force microscopy or mechanical manipulation,[1, 2] where individual elements are physically relocated using the scanning tip. Other techniques such as spin coating and solution dipping[3-5] result in random positioning of emitters relative to waveguides and often do not give rise to any optical enhancement.

An alternative, promising approach for the positioning of nanophotonic elements is provided by bottom-up self-assembly techniques.[6] Indeed, wet chemistry has been used to assemble one-dimensional chain-like structures by ordering gold and silver nanoparticles of different shapes, including spheres,[7] cubes,[8, 9] and rods.[9, 10] Inspired by this approach, we introduce a simple, fast and cost effective chemical self-assembly method for the attachment of two primary components of a practical plasmonic circuit: a single photon emitter and a waveguide. Our method enables coupling of nanodiamonds with a single quantum emitter (the nitrogen-vacancy (NV) center) onto the terminal of a silver nanowire. We chose these particular elements because the NV center acts as a room temperature quantum emitter with a plethora of applications in sensing[11] and quantum information,[12] while silver nanowires act as metallic waveguides that support surface plasmon polariton (SPP) propagation modes.

The chemical self-assembly procedure is depicted in Figure 1a. As-synthesized silver nanowires[13] fluorescent nanodiamonds  ascorbic acid (AA) and deionized water are mixed in a vial, ultrasonicated for 10 min and a portion of the solution is drop-cast onto a $SiO_2$ substrate (detailed descriptions of silver nanowire synthesis and preparation of nanodiamonds are provided in the Supporting Information). The as-synthesized nanowires are coated with polyvinylpyrrolidone (PVP).[13] To control the density of nanodiamonds attached to the nanowires, the concentration of AA is varied. Figure 1(b-d) shows typical hybridized nanowire/nanodiamonds nanostructures fabricated using AA concentrations of 25, 5 and 1mM, respectively. At high AA concentrations, nanodiamonds assemble randomly over the bodies of the silver nanowires. Conversely, at low AA concentration (1 mM), single nanodiamonds bind preferentially onto the terminals of the silver nanowires. In order to verify the effectiveness of the self- assembly process, we collect low-magnification TEM images of several tens of randomly chosen silver nanowire – nanodiamond(s) hybrid structures (Figure S2). Hybrid structures with nanodiamond(s) attached at the apex of each nanowire account for roughly more than 50% of the investigated population. The upper panels of Figure 1(b-d) show representative low magnification SEM images and corresponding schematic illustrations of the hybrid systems. Energy-dispersive x-ray (EDX) spectral mapping (Figure 1e) was used to determine the elemental composition of the hybrid nanostructures. The analysis confirmed that the features seen in SEM images are silver nanowires (green) and carbon-rich particles (red).

To confirm that the particles are diamonds, selected-area electron diffraction patterns were obtained from the body and the apex of a silver nanowire (Figure S3). The first pattern is characteristic of a pentagonally-twinned silver nanowire, consistent with prior analyses of silver nanowires,[14] and gold nanorods.[15] The second diffraction pattern was obtained from an area that includes both the apex of the nanowire and an attached nanoparticle. It was indexed as the combination of a pentagonally-twinned silver nanowire and diamond.

High resolution transmission electron microscopy (HRTEM) was used to elucidate the role of PVP and AA in the attachment of nanodiamonds to the silver nanowires. The image in Figure 2a shows a typical interface between a nanodiamond and a silver nanowire. The nanodiamond is attached directly to the surface of the silver nanowire with no obvious amorphous layers in between (we note, however, that we do not expect ascorbate molecules to be identifiable in the TEM data and hence can not eliminate the possibility of their incorporation at the interface). Figure 2b shows a nanowire region that does not contain a nanodiamond, but contains an amorphous layer that we attribute to PVP. As-synthesized nanowires are known to be terminated by a layer of PVP,[14, 16, 17] consistent with our TEM data. However, importantly, the PVP molecules accumulate preferentially away from the nanowire apexes as seen in the insets of Figure 2c (and Figure S4a) where the thickness of the amorphous layer seen at the apex is approximately half that along the body. This observation is consistent with previous investigations that suggest preferential adsorption of PVP molecules to (100) facets over (111) facets of silver nanowires synthesized by the polyol route.[16] Since the surface free energy of PVP layers[18] (~47 mJ/m$^2$) is significantly lower than that of the silver NW surface[19] (~890 mJ/m$^2$), the PVP serves to passivate the highly active silver NWs. However, addition of AA to the solution causes the PVP molecules to aggregate into clusters (see Figure S4b), thereby exposing and activating adjacent regions of the NW surface. The apexes are activated preferentially due to the smaller quantity of PVP molecules at the tips of as-grown nanowires. Hence, the nanodiamonds are preferentially attached to the apexes, provided the concentration of AA is sufficiently low to avoid exposure of surface regions along the entire length of the nanowire. At high AA concentrations, the entire nanowires contain exposed surface regions (Figure S4c), promoting random attachment of nanodiamonds along the entire length of each nanowire.

AA can therefore be used to control the locations and concentration of nanodiamonds attached to the silver NWs. To demonstrate this effect, the assembly procedure was repeated using different concentrations of AA (the results of a control experiment performed without nanodiamonds are shown in Figure S4). In the absence of AA, the nanowires are covered with PVP and do not contain attached nanodiamonds (Figure 2c). A low concentration of AA (1 mM) yields nanowires with nanodiamonds attached preferentially to their apexes (Figure 2d). A higher concentration of AA (10 mM) results in random attachment of nanodiamonds along the entire surface of the silver nanowires (Figure 2e), while an AA concentration of 50 mM yields a high concentration of large amorphous clusters (PVP agglomerates) attached to the nanowire surface (Figure 2f).

We therefore conclude that AA causes the agglomeration of PVP molecules into clusters, as shown schematically in Figure 2g. By measuring pH values, we also observe that there is a dramatic reduction in pH caused by the introduction of AA (Figure S5). At low AA concentrations, the nanowire apexes are exposed preferentially because the thickness of the PVP layer is smallest at the tips of the as-grown nanowires, resulting in selective attachment of nanodiamonds to these regions. High AA concentrations cause the exposure of surface regions over the entire length of the silver nanowires, leading to random attachment of nanodiamonds to the nanowires.

We now examine the performance of the fabricated silver nanowire/nanodiamonds hybrids as devices for use in nanophotonics. Quantum optical characterization is used to quantify the coupling efficiency of NV centers to the plasmonic mode of a silver nanowire. A confocal microscope with a Hanbury Brown and Twiss interferometer was used to record the second-order autocorrelation functions, $g^{(2)}(\tau)$, shown in Figure 3a from a coupled and an uncoupled NV center using a 532 nm excitation laser. The dips below 0.5 at zero delay time ($g^{(2)}(0) < 0.5$) seen in the figure are characteristic of single photon emitters.[20, 21] The data were fit using a three-level model[20] of the NV center:

$$g^{(2)}(\tau) = 1 + c_2 e^{-\tau/\tau_2} + c_3 e^{-\tau/\tau_3} \quad (1)$$

where $\tau_2$ and $\tau_3$ are the lifetime of the excited and shelving states, respectively.

Fluorescence lifetime measurements (Figure 3b) fit using double-exponential functions[22] confirm that the coupled NV center possesses a significant reduction in lifetime ($\tau = 3.15$ ns) compared to the uncoupled NV center ($\tau = 11.98$ ns). On average, an enhancement factor of 2.44 was obtained which is in good agreement with prior theoretical predictions.[3, 4] Figure S6 shows the statistical distribution of the lifetime reduction of the coupled NV centers. The observed variation is expected due to a corresponding variation in the orientation of the NV center dipole with respect to the silver nanowire. To complete the optical characterization, saturation measurements were obtained from the coupled and uncoupled NV centers (Figure 3c). As expected, the coupled NV center exhibits a higher fluorescence efficiency at each laser power. The data were fit using the three level saturation model:[20]

$$I = \frac{I_\infty P}{P_{sat}+P} \quad (2)$$

which was used to obtain the saturation fluorescence intensity $I_\infty$ and the corresponding laser power $P_{sat}$. Saturation laser powers of 325 and 422 µW, which correspond to fluorescence intensities of ~ 102 and 47 kcounts/s, were obtained for the coupled and uncoupled NV centers, respectively. This saturation behavior and the average count rates are consistent with previous studies.[20, 23] The observed reduction in lifetime associated with the increase in fluorescence intensity unambiguously demonstrate plasmonic enhancement in the hybrid system.[2]

Finally, we study SPP propagation in the self-assembled structures using the confocal microscope configuration with two collection channels shown in Figure 4a. Plasmon propagation through the nanowire is demonstrated in Figure 4b. The laser was fixed at point A, exciting a single NV center coupled to a silver nanowire, while a confocal map was collected by scanning the second collection channel in a raster pattern. Fluorescence emission from the NV center into free-space was detected at point A (brighter spot), while out-coupling of the SPP mode was observed at the other end of a silver nanowire (point B). The lower fluorescence intensity of point B is attributed to ohmic loss of the SPP along the silver nanowire.[3] An SEM image of the silver nanowire is shown in Figure 4b to emphasize the propagation direction. Figure 4c shows fluorescence spectra taken from points

A and B. Spectrum A is a typical NV spectrum, while spectrum B is distorted due to silver fluorescence and the self-interference behavior reported elsewhere.[4]

To summarize, we have presented a facile and cost effective self-assembly method for efficient coupling of single quantum emitters to metallic nanowires. The method is based on a chemical process in which AA is used to activate the apexes of PVP-terminated nanowires. The quantity of the attached nanodiamonds can be controlled by varying the concentration of AA. The proof of principle quantum hybrid system presented in this work will serve as a robust building block for the fabrication of an integrated quantum plasmonic network with multiple elements. The method can be adapted for use with other systems such as dielectric optical cavities, plasmonic antennas and metamaterials.

**Acknowledgements**
We thank Brendan Shields for assistance with the assembly of the dual arm confocal microscope. The work was supported in part by the Australian Research Council (Project Number DP140102721) and FEI Company. I. A. is the recipient of an Australian Research Council Discovery Early Career Research Award (Project Number DE130100592). O.S. acknowledges the Ramaciotti Foundation for the financial support.

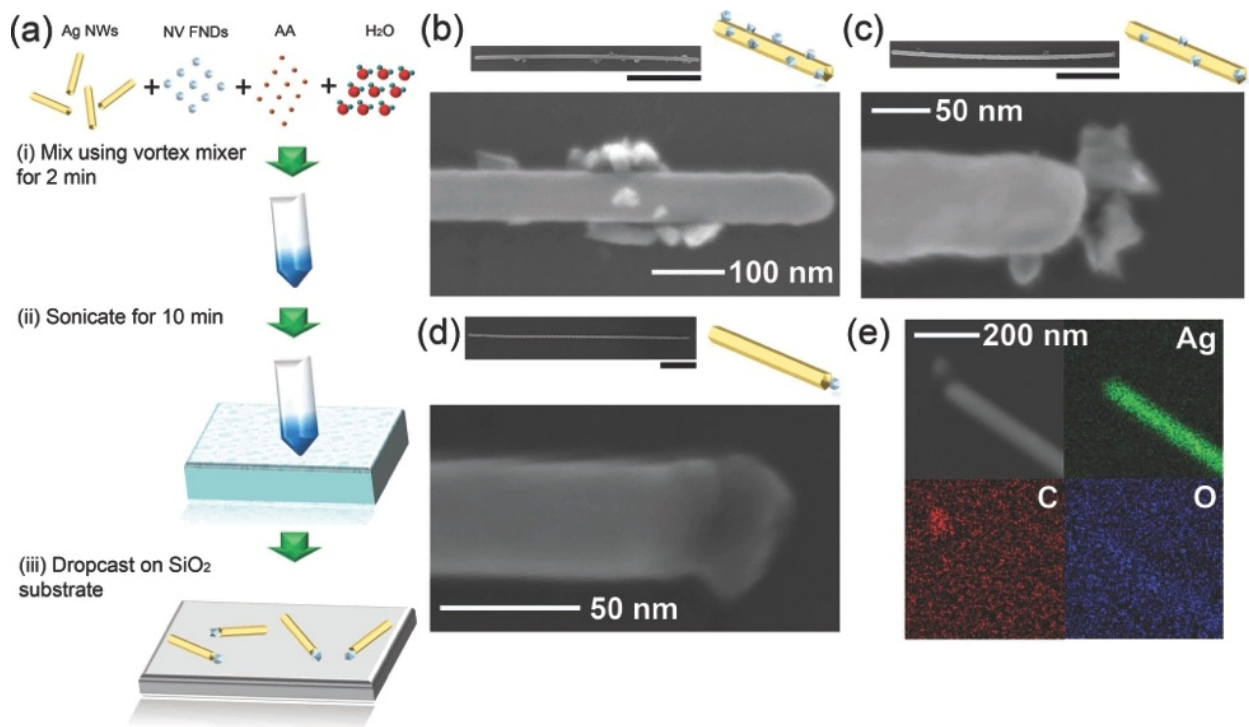

**Figure 1. Selective attachment of nanodiamonds to the apexes of silver nanowires.** (a) Schematic illustration of the self-assembly process: (i) silver nanowires, nanodiamonds, ascorbic acid and deionized water are mixed using a vortex mixer for 2 min, (ii) the solution is sonicated for 10 min, and (iii) a small portion of the solution is drop-cast onto a SiO$_2$ substrate and dried overnight at room temperature. (b-d) SEM images and schematic illustrations of typical silver nanowire/nanodiamond hybrids fabricated using AA concentrations of: (b) 25 mM, (c) 5 mM, and (d) 1 mM. The scale bar of each low-magnification SEM image is 1 μm. (e) SEM image and EDX maps of a typical silver nanowire/nanodiamond hybrid obtained using an AA concentration of 1 mM.

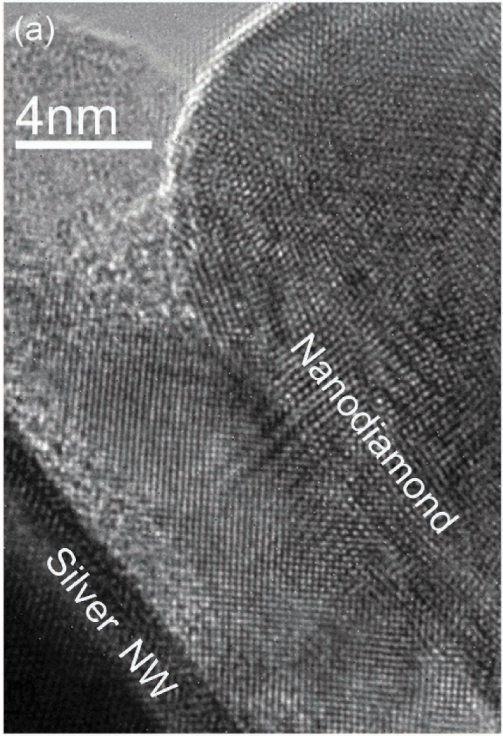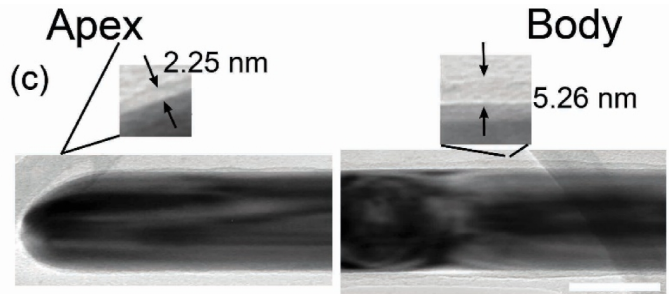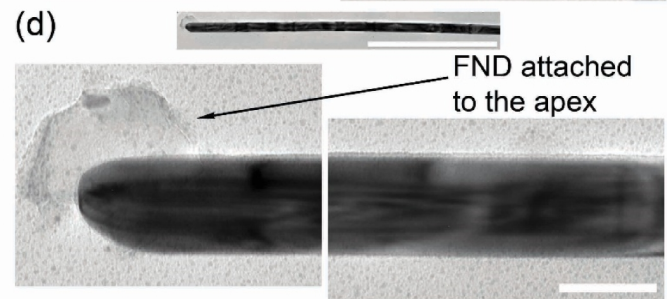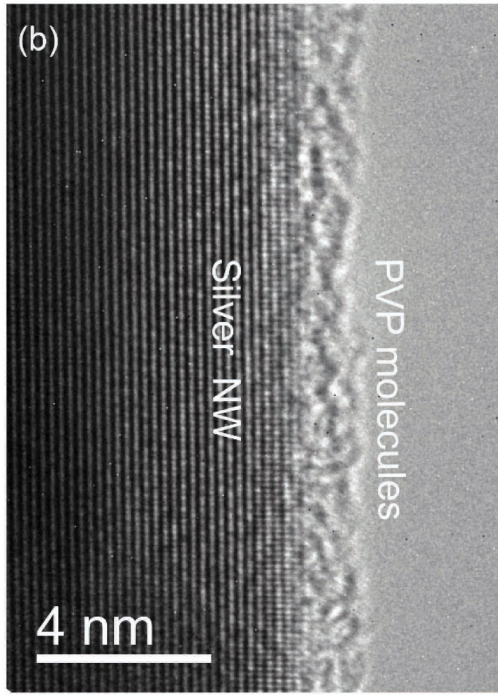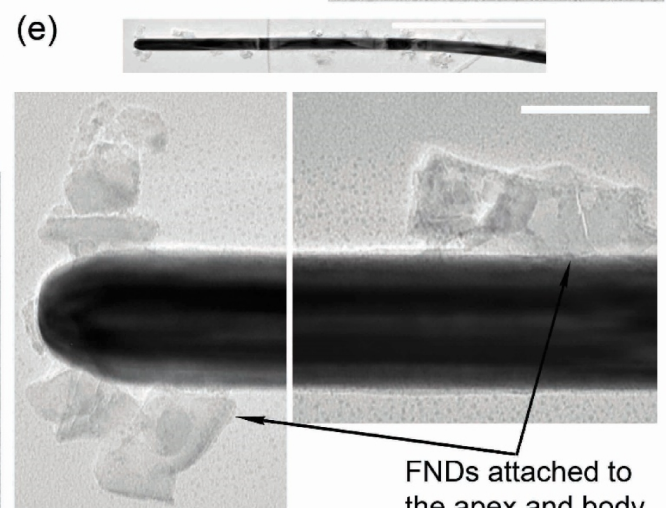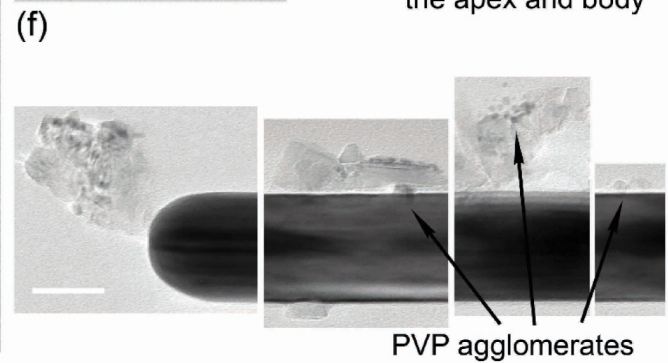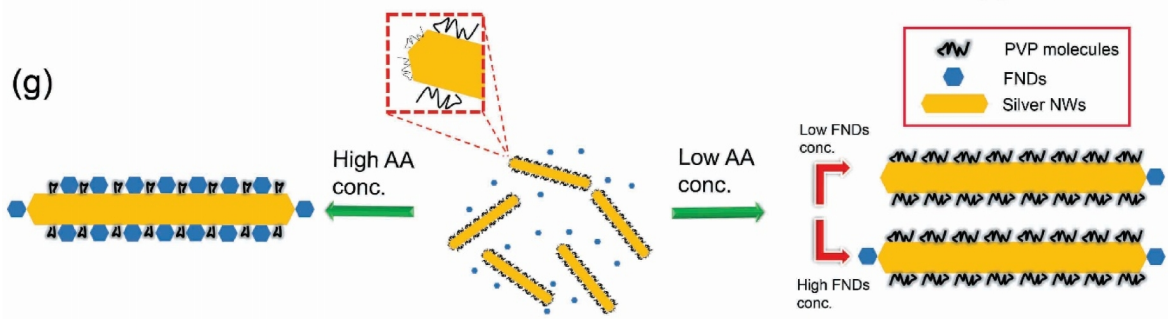

**Figure 2. TEM characterization of silver nanowire/nanodiamonds hybrids.** (a) Typical interface between a silver nanowire and an attached nanodiamond. The lattice fringes of the nanodiamond are clearly visible and the nanodiamond is attached to the silver nanowire without an intermediate organic layer. (b) A layer of PVP molecules on the surface of a silver nanowire, that is formed during nanowire synthesis. (c-f) Typical assemblies of silver nanowires and nanodiamonds fabricated using AA concentrations of (c) 0 mM, (d) 1 mM, (e) 10 mM, and (f) 50 mM. The scale bar on each image and the corresponding inset represents 50 nm and 1μm, respectively. (g) Schematic illustration of the effect of AA concentration on the attachment of nanodiamonds to silver nanowires. Selective attachment to nanowire tips occurs at low AA concentration.

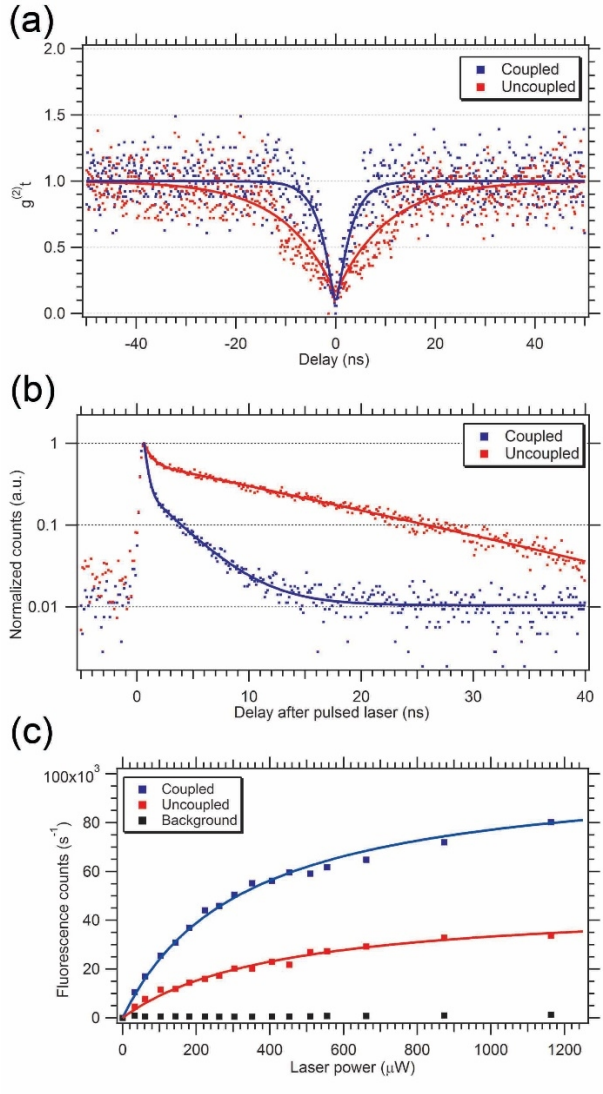

**Figure 3. Optical characterization of a NV center coupled to a silver nanowire:** (a) second-order autocorrelation function, g$^{(2)}$(τ), recorded from a coupled (blue squares) and an uncoupled (red squares) nanodiamond. The dip at zero delay time indicates that a single emitter (NV center) is addressed. The solid lines are fits to the data generated using a three level model. (b) Fluorescence lifetime data of a single coupled (blue squares) and an uncoupled (red squares) nanodiamond. The solid lines are fits to the data produced using a double exponential decay model. (c) Fluorescence intensity as a function of excitation power obtained from a single NV center coupled to a nanowire (blue squares) and an uncoupled NV center (red squares). Black squares

indicate the background collected from the substrate. Solid lines are fits to the experimental data generated using a three level saturation model. A laser power of 400 µW and a 697 ± 37.5 nm band-pass filter were used to perform all measurements.

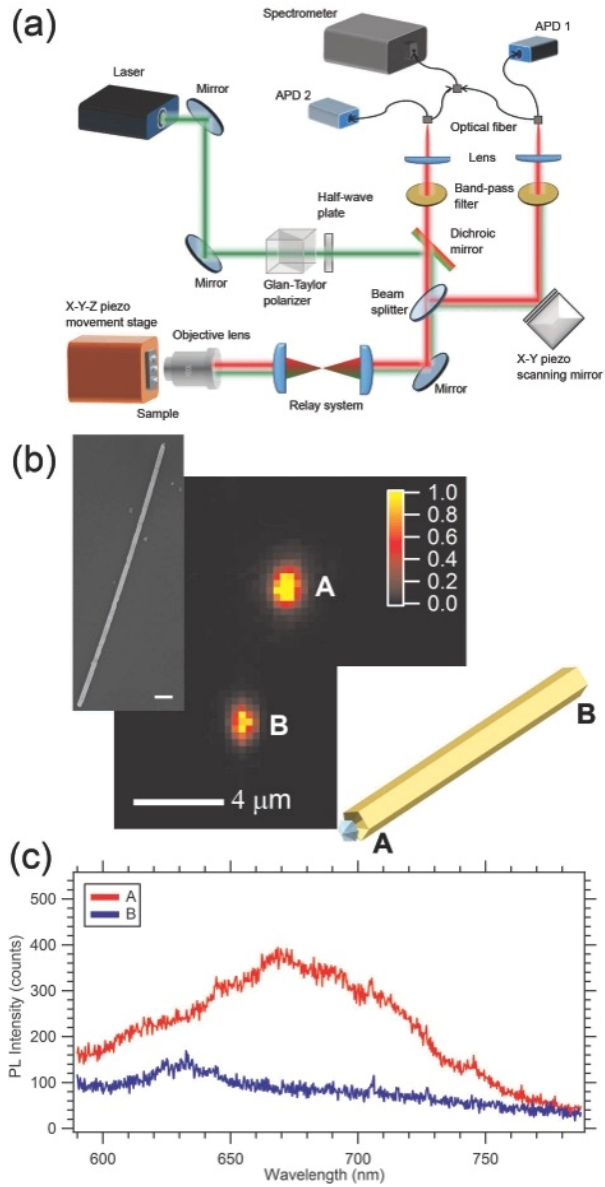

**Figure 4. Plasmonic coupling between a single NV center and a silver nanowire.** (a) The experimental set-up in which the excitation spot is decoupled from the collection spot using an additional X-Y piezo scanning mirror [APD = avalanche photo diode]. (b) A confocal PL map obtained using an excitation laser fixed at a single NV center (point A). Emission at point B is detected using the second confocal arm and indicates the out-coupled emission that propagated along the nanowire. The inset shows a representative SEM image of a silver

nanowire/nanodiamond hybrid (the scale bar is 300 nm). (c) PL spectra collected from points A (red curve) and B (blue curve) of the confocal map shown in (b).